\documentclass[12pt]{article}

\usepackage{times}

\DeclareFontFamily{OT1}{times}{}
\DeclareFontShape{OT1}{times}{m}{n }{ <-> ptmr }{}
\DeclareFontShape {OT1}{times}{bx}{n }{ <-> ptmb }{}
\DeclareFontShape {OT1}{times}{m }{it}{ <-> ptmri}{}
\DeclareFontShape {OT1}{times}{bx}{it}{ <-> ptmbi}{}
\usepackage{latexsym}
\usepackage{graphics}

\begin{document}

\newtheorem{theo}{Theorem}[section]
\newtheorem{defi}[theo]{Definition}
\newtheorem{prop}[theo]{Proposition}
\newtheorem{corr}[theo]{Corollary}
\newtheorem{lemm}[theo]{Lemma}
\newtheorem{exam}[theo]{Example}

\newcommand{\CD}[2]{\ensuremath{\raisebox{-.6ex}{\scriptsize{$0$}}\:\!
        \mathbf{C}_{t}^{#1}\left[#2\right]}}
\newcommand{\FFD}[4]{\ensuremath{\raisebox{-.6ex}{\scriptsize{$#1$}}\!
        \raisebox{1ex}{\scriptsize{$t$}}\:\!\mathbf{F}_{#2}^{#3}[#4]}}
\newcommand{\myH}[1]{\ensuremath{H^{#1}([0,T])}}
\newcommand{\K}[3]{\ensuremath{K_{#1}^{(#2)}(#3)}}
\newcommand{\Kstar}[3]{\ensuremath{K_{#1}^{(#2)} \star q^{#3}}}
\newcommand{\myL}{\ensuremath{L^{2}([0,T])}}
\newcommand{\LFD}[3]{\ensuremath{\raisebox{-.6ex}{\scriptsize{$#1$}}\:\!
        \mathbf{D}_{t}^{#2}\left[#3\right]}}
\newcommand{\LFDm}[3]{\ensuremath{\raisebox{-.6ex}{\scriptsize{$#1$}}\:\!
        \mathbf{D}_{t^{-}}^{#2}\left[#3\right]}}
\newcommand{\LFI}[3]{\ensuremath{\raisebox{-.6ex}{\scriptsize{$#1$}}\:\!
        \mathbf{I}_{t}^{#2}[#3]}}
\newcommand{\mysum}{\ensuremath{\sum_{n=1}^{\infty}}}
\newcommand{\PHIm}[2]{\ensuremath{\Phi_{#1}^{-}(#2)}}
\newcommand{\PHIp}[2]{\ensuremath{\Phi_{#1}^{+}(#2)}}
\newcommand{\PSI}[2]{\ensuremath{\Psi_{#1}(#2)}}
\newcommand{\lfeta}[2]{\ensuremath{\raisebox{-.6ex}{\scriptsize{$#1$}}\:\!\eta_{t}^{#2}}}
\newcommand{\rfeta}[2]{\ensuremath{\raisebox{-.6ex}{\scriptsize{$t$}}\:\!\eta_{#1}^{#2}}}
\newcommand{\lfq}[2]{\ensuremath{\raisebox{-.6ex}{\scriptsize{$#1$}}\:\!q_{t}^{#2}}}
\newcommand{\RFD}[3]{\ensuremath{\raisebox{-.6ex}{\scriptsize{$t$}}\:\!
        \mathbf{D}_{#1}^{#2}\left[#3\right]}}
\newcommand{\RFDp}[3]{\ensuremath{\raisebox{-.6ex}{\scriptsize{$t^{+}$}}\:\!
        \mathbf{D}_{#1}^{#2}\left[#3\right]}}
\newcommand{\RFI}[3]{\ensuremath{\raisebox{-.6ex}{\scriptsize{$t$}}\:\!
        \mathbf{I}_{$#1$}^{#2}[#3]}}
\newcommand{\RLD}[2]{\ensuremath{\raisebox{-.6ex}{\scriptsize{$0$}}\:\!
        \mathbf{R}_{t}^{#1}\left[#2\right]}}
\newcommand{\SIP}[3]{\ensuremath{\langle q_{#1},q_{#2}\rangle_{H^{#3}}}}
\newcommand{\Test}{\ensuremath{C^{\infty}([0,T])}}


\title{Nonconservative Lagrangian Mechanics: A generalized function approach}

\author{David W. Dreisigmeyer\footnote{email:davidd@engr.colostate.edu}\ \ and
        Peter M. Young\footnote{email:pmy@engr.colostate.edu}\\
        Department of Electrical and Computer
        Engineering\\
        Colorado State University, Fort Collins, CO 80523\\}

\maketitle
\begin{abstract}
We reexamine the problem of having nonconservative equations of
motion arise from the use of a variational principle.  In
particular, a formalism is developed that allows the inclusion of
fractional derivatives.  This is done within the Lagrangian
framework by treating the action as a Volterra series.  It is then
possible to derive two equations of motion, one of these is an
advanced equation and the other is retarded.
\end{abstract}

\section{Introduction}
\label{SEC-INTRO}

The problem of having a dissipation term $\dot{q}$ arise in the
equations of motion for a system has a long history.  Bauer
\cite{Baue31} showed that ``the equations of motion of a
dissipative linear dynamical system with constant coefficients are
not given by a variational principle''.  There are loopholes in
Bauer's proof, however.  One of these is to allow for additional
equations of motion to arise.  This method was employed by Bateman
\cite{Bate31}.  He used the Lagrangian
\begin{eqnarray}\label{Intro1}
L & = & m\dot{x}\dot{y} +\frac{C}{2}\ (x\dot{y} -\dot{x} y)
\end{eqnarray}
which gives the equations of motion
\begin{eqnarray}\label{Intro2}
\begin{array}{ccccccc}
m\ddot{x} + C \dot{x} & = & 0 & & m\ddot{y} - C\dot{y} & = & 0
\end{array}
\end{eqnarray}
Bateman's method is not very general, so we look for other methods
to model nonconservative systems.

Caldeira and Leggett \cite{CaLe83} suggest recognizing that a
dissipative system is coupled to an environment.  The environment
is modelled as a collection of harmonic oscillators which results
in the Lagrangian
\begin{eqnarray}\label{Intro3}
L & = & \frac{m}{2}\ \dot{q}^{2} - V(q) + \sum_{n=1}^{\infty}
        \left\{ \frac{m_{n}}{2}\ \dot{q}_{n}^{2} - \frac{m_{n}
        \omega_{n}^{2}}{2}\ (q_{n}-q)^{2} \right\}
\end{eqnarray}
where $q$ is the system's coordinate and the $q_{n}$'s are the
environment's coordinates.  While the system by itself is
nonconservative, the system plus environment is conservative.
This procedure does allow the introduction of very general
dissipation terms into the system's equation of motion. However,
the microscopic modelling of the environment makes (\ref{Intro3})
much more complex than, say, (\ref{Intro1}).

In order to overcome the difficulties of the above two procedures,
Riewe examined using fractional derivatives in the Lagrangians
\cite{Riew96,Riew97}.  This method takes advantage of another
loophole in Bauer's proof.  Namely, Bauer assumed that all
derivatives were integer ordered.  Riewe's method has the
advantage of not introducing extra coordinates as in
(\ref{Intro1}) and (\ref{Intro3}).  However, it ultimately results
in noncausal equations of motion.  A rather ad hoc procedure of
replacing anti-causal with causal operators needs to be used at
the end in order to arrive at causal equations of motion.  We will
present a method that can be used within Riewe's formalism that
avoids this situation.

We propose here a new method of using a variational principle to
derive nonconservative equations of motion.  Our method is closely
related to Riewe's in that we use fractional operators. However,
we treat these operators as kernels in a Volterra series. We show
that Riewe's formalism can be derived by using certain types of
symmetric kernels in the series expansion.  A simple modification
of the kernels will result in two equations of motion for a
system. One of these equations is advanced while the other is
retarded, similar to (\ref{Intro2}).

Our paper is organized as follows.  In Section \ref{SEC-FRAC} we
review fractional integration and differentiation.  Riewe's
formalism is briefly examined in Section \ref{SEC-Riewe}.  We then
give a brief overview of Volterra series in Section
\ref{SEC-Volterra} before examining our fractional Lagrangian
mechanics in Section \ref{SEC-FRACMECH}.  Section \ref{SEC-ENVIR}
examines the nonconservative harmonic oscillator in a different
way than the traditional variational methods.  A discussion of
some related concepts and future research follows in Section
\ref{SEC-Discussion}.

\section{Fractional Integration and Differentiation}
\label{SEC-FRAC}

Fractional integrals and derivatives are generalizations of their
usual integer ordered operations.  To start developing the theory,
let us first write down Cauchy's integral formula
\begin{eqnarray}\label{Cauchy}
f^{(-n)}(t) & = & \frac{1}{\Gamma(n)}\ \int_{a}^{t} f(\tau)
        (t-\tau)^{n-1} d\tau
\end{eqnarray}
where $n > 0$ is an integer, $\Gamma(n)$ is the gamma function,
and $a<t$.  Equation (\ref{Cauchy}) is a convolution of $f(t)$ and
the function
\begin{eqnarray}\label{PHIplusdef}
\PHIp{n}{t} & := & \left\{ \begin{array}{cc}
            \frac{1}{\Gamma(n)}\ t^{n-1} & t>0  \\
            0 & t \leq 0
            \end{array}\right.
\end{eqnarray}
if we set $f(t) \equiv 0$ for $t < a$.  So we can rewrite
(\ref{Cauchy}) as
\begin{eqnarray}\label{LFI1}
\LFI{a}{n}{f} & = & f(t) \ast \PHIp{n}{t}
\end{eqnarray}
where $\ast$ is the convolution operation defined by
\begin{eqnarray}\label{convolution}
g(t) \ast h(t) & := & \int_{-\infty}^{\infty} g(\tau) h(t-\tau)
            d\tau
\end{eqnarray}
Equation (\ref{LFI1}) will be our stepping stone to generalizing
the integer ordered operations to fractional order.

The above procedure works so well for the integers $n>0$, we want
to consider extending it to any real $\alpha > 0$.  This is
obviously possible, so we let
\begin{eqnarray}\label{LFI}
\LFI{a}{\alpha}{f} & = & f(t) \ast \PHIp{\alpha}{t}
\end{eqnarray}
be the left fractional integral [LFI] of $f(t)$ of order $\alpha >
0$.  Everything works fine until we consider the case $\alpha =
0$.  We reasonably expect that
\begin{eqnarray}\label{LFI0}
\LFI{a}{0}{f} & = & f(t)
\end{eqnarray}
but, it is not immediately obvious that the integral in
(\ref{LFI1}) is not divergent.  Also, for $-1 < \alpha < 0$, the
integral is obviously divergent.  It is apparent that treating
$f(t)$ and $\PHIp{\alpha}{t}$ as regular functions will not be
sufficiently general for our purposes.  Instead we will consider
them to be distributions, or generalized functions.  [We note that
there are other ways to generalize integer ordered derivatives to
fractional order \cite{Podl99}.  We will work solely with the
generalized function approach to interpolate between the integer
ordered integrals and derivatives.]

The first order of business is to define the convolution operation
for distributions.  Let $k(t) = g(t) \ast h(t)$ and $\varphi(t)$
be a test function.  Then \cite{GeSh64}
\begin{eqnarray}\label{convgf}
\langle k,\varphi\rangle & := & \int k(t) \varphi(t) dt   \nonumber\\
        & = & \int \left\{ \int g(\xi) h(t - \xi) d\xi \right\}
                \varphi(t) dt          \\
        & = & \int \int g(\xi) h(\eta) \varphi(\xi + \eta) d\xi d\eta
                    \nonumber
\end{eqnarray}
Equation (\ref{convgf}) is meaningful as long as either $g(t)$ or
$h(t)$ has bounded support or, $g(t)$ and $h(t)$ are bounded on
the same side [e.g., $g(t) \equiv 0$ for $t<t_{1}$ and $h(t)
\equiv 0$ for $t<t_{2}$].  We will always assume that one of these
situations is the case.  From (\ref{convgf}), it can be seen that
the generalization of (\ref{convolution}) is
\begin{eqnarray}\label{GFconv}
\langle g \ast h,\varphi\rangle & = & \langle g(t), \langle
        h(\tau),\varphi(t+\tau) \rangle\rangle
\end{eqnarray}
The convolution operation has the properties
\begin{eqnarray}
g\ast h & = & h \ast g  \label{GFconvP1}    \\
f\ast (g\ast h) & = & (f\ast g)\ast h    \label{GFconvP2}    \\
D(g\ast h) & = & (Dg)\ast h  =  g\ast(Dh) \label{GFconvP3}
\end{eqnarray}
where $D(\cdot)$ is the generalized derivative.  Remember that the
relationship between the generalized and classical derivatives,
beginning at $t=a$, is given by \cite{Podl99}
\begin{eqnarray}\label{CGrel}
D^{n} f & = & f^{(n)} + \sum_{k=0}^{n-1} \left[
            D^{n-k-1}\delta(t-a) \right]
            f^{(k)}(a)
\end{eqnarray}
where $f^{(n)}$ is the classical derivative.

Considering $\PHIp{\alpha}{t}$ as a generalized function allows us
to extend (\ref{LFI}) to any $\alpha$, where the convolution
operation is defined as in (\ref{GFconv}).  For $\alpha < 0$, this
will define the left fractional derivative [LFD] as
\begin{eqnarray}\label{LFD}
\LFD{a}{-\alpha}{f} & := & \LFI{a}{\alpha}{f}   \nonumber   \\
        & = & f(t) \ast \PHIp{\alpha}{t}
\end{eqnarray}
In the sequel, we will find it easier to assume $\alpha > 0$ and
use the notation
\begin{eqnarray}\label{LFDI}
\begin{array}{ccccccc}
\LFD{a}{\alpha}{f} & = & f(t)\ast \PHIp{-\alpha}{t} & &
    \LFD{a}{-\alpha}{f} & = & \LFI{a}{\alpha}{f}
\end{array}
\end{eqnarray}
Also, for reasons that will become apparent shortly, we will often
set $f(t) \equiv 0$ for $t<a$ and $t>b$, where $a<b$.  We do not
want any resulting discontinuities in $f(t)$ at $t = b$ to affect
the LFDs.  So $t$ must be restricted to the interval $a \leq t <b$
in the LFDs.  It would perhaps be better to write (\ref{LFDI}) as
\begin{eqnarray}\label{LFD-}
\LFDm{a^{-}}{\alpha}{f} & = & \frac{1}{\Gamma(-\alpha)}\
    \int_{a^{-}}^{t^{-}} f(\tau) (t-\tau)^{-(\alpha+1)} d\tau
\end{eqnarray}
To avoid cluttering our notation, we will continue to use the
notation in (\ref{LFDI}) with the understanding that it formally
means (\ref{LFD-}).

The distributions $\PHIp{\alpha}{t}$ have been well studied
\cite{GeSh64,Podl99}.  Their two most important properties are
\begin{eqnarray}\label{deltader}
\PHIp{n}{t} & = & D^{-n}\delta(t^{+})
\end{eqnarray}
for any integer $n$, and, for any $\beta$ and $\gamma$,
\begin{eqnarray}\label{phiadd}
\PHIp{\beta}{t-a} \ast \PHIp{\gamma}{t} & = & \PHIp{\beta +
        \gamma}{t-a}
\end{eqnarray}
Equation (\ref{phiadd}) implies
\begin{eqnarray}
\LFD{a}{\beta}{\LFD{a}{\gamma}{f}} & = & \LFD{a}{\beta +
        \gamma}{f}  \label{LFDadd}  \\
\LFD{a}{\beta}{\LFD{a}{-\beta}{f}} & = & f  \label{LFDinv}
\end{eqnarray}

Now let $0 \leq n -1 \leq \alpha < n$.  Then, using
(\ref{GFconvP1}) -- (\ref{GFconvP3}) and (\ref{deltader}) and
(\ref{phiadd}), we have
\begin{eqnarray}
\LFD{a}{\alpha}{f} & = & f(t) \ast \PHIp{-\alpha}{t}
        \nonumber               \\
        & = & f(t) \ast \left(D^{n} \PHIp{n-\alpha}{t}\right)
        \nonumber   \\
        & = & \left(D^{n} f(t)\right) \ast \PHIp{n-\alpha}{t}
        \label{Caputo}  \\
        & = & D^{n}\left( f(t) \ast \PHIp{n-\alpha}{t}\right)
        \label{RLder}
\end{eqnarray}
Equations (\ref{Caputo}) and (\ref{RLder}) are the distributional
forms of the Caputo and Riemann-Liouville fractional derivative,
respectively \cite{Podl99}.  In the standard definitions of these
derivatives, $D^{n}$ is replaced with $(d/dt)^{n}$.

In addition to the left fractional operations, we can also define
right fractional operations.  If we set $f(t) \equiv 0$ for $t >
b$ and define
\begin{eqnarray}\label{PHIminusdef}
\PHIm{\alpha}{t} & := & \left\{ \begin{array}{cc}
        \frac{1}{\Gamma(\alpha)}\ (-t)^{\alpha-1} & t<0 \\
        0 & t \geq 0
        \end{array}\right.
\end{eqnarray}
the right fractional operations are defined by
\begin{eqnarray}\label{RFD}
\RFD{b}{\alpha}{f} & := & f(t) \ast \PHIm{-\alpha}{t}
\end{eqnarray}
Most of the above observations for the left fractional operations
also hold for the right ones.  However, (\ref{deltader}) needs to
be replaced with
\begin{eqnarray}\label{RFDprop1}
\PHIm{n}{t} & = & (-1)^{n} D^{-n} \delta(t^{-})
\end{eqnarray}
for any integer $n$.  When $f(t) \equiv 0$ for $t<a$ and $t>b$, we
do not allow any resulting discontinuities in $f(t)$ at $t=a$ to
affect the RFDs.  Similar to the case for the LFDs, we will take
(\ref{RFD}) as meaning
\begin{eqnarray}\label{RFD+}
\RFDp{b^{+}}{\alpha}{f} & = & \frac{1}{\Gamma(-\alpha)}\
    \int_{t^{+}}^{b^{+}} f(\tau) (\tau - t)^{-(\alpha+1)}
\end{eqnarray}
though we will continue to use the notation in (\ref{RFD}).

Note that for the left operations, the ``left'' integration limit
$a$ determines the allowable functions in the operation
$\LFD{a}{\alpha}{f}$.  Namely, $f(t)$ must vanish for $t<a$. Also,
$\LFD{a}{\alpha}{f}$ is a function of $\alpha$ and $t$ and, a
functional of $f(t)$.  Similar comments hold for the right
operations.  Here, the ``right'' integration limit $b$ means $f(t)
\equiv 0$ for $t>b$.  Now let $f(t)$ be compactly supported on the
interval $[a,b]$.  Then $\LFD{a}{\alpha}{f} = 0$ whenever $t<a$.
However, $\LFD{a}{\alpha}{f}$ does not generally vanish for $t>a$.
Thus, the left operations are causal or retarded.  Conversely,
$\RFD{b}{\alpha}{f} = 0$ whenever $t>b$ but, generally,
$\RFD{b}{\alpha}{f} \neq 0$ for $t<b$.  Hence, the right
operations are anti-causal or advanced.

Our fractional derivatives satisfy an integration by parts
formula.  First, assume that $f(t) \equiv 0$ for $t<a$ and $g(t)
\equiv 0$ for $t>b$.  Then, for any $\beta$,
\begin{eqnarray}\label{IBP1}
\langle g(\tau) \PHIp{\beta}{\tau-t} f(t), \varphi(t,\tau) \rangle
        & = & \langle g(\tau) \PHIm{\beta}{t-\tau} f(t),
        \varphi(t,\tau)\rangle
\end{eqnarray}
Hence,
\begin{eqnarray}\label{IBP2}
\langle g (\Phi_{\beta}^{+} \ast f ), \varphi \rangle
        &=& \langle (g \ast \Phi_{\beta}^{-}) f, \varphi \rangle
\end{eqnarray}
or
\begin{eqnarray}\label{IBP}
\int \LFD{a}{\beta}{f} g(t) dt & = & \int \RFD{b}{\beta}{g} f(t)
        dt
\end{eqnarray}
We note that Riewe's derivation of an integration by parts formula
\cite[Equation (16)]{Riew97} is flawed on two points.  First, the
boundary conditions are generally fractional, not integer,
ordered.  Also, Riewe incorrectly exchanges the classical Caputo
derivative [(\ref{Caputo}) with $D^{n}$ replaced with
$(d/dt)^{n}$] for the Riemann-Liouville derivative in
(\ref{RLder}). Fortunately, when vanishing boundary conditions are
assumed, these defects are inconsequential.  Also notice that
(\ref{IBP}) implies that any integration by parts inherently
introduces time reversal.

When we examine Riewe's fractional mechanics in Section
\ref{SEC-Riewe}, (\ref{IBP}) will lead to equations of the form
\begin{eqnarray}\label{dder}
\Phi_{\beta}^{-} \ast \left( \Phi_{\beta}^{+} \ast f \right) & = &
        \RFD{b}{\alpha}{\LFD{a}{\alpha}{f}} \nonumber   \\
        &\stackrel{?}{=} & \left( \Phi_{\beta}^{-} \ast
        \Phi_{\beta}^{+} \right) \ast f
\end{eqnarray}
The difficulty with (\ref{dder}) is that neither are
$\PHIp{\beta}{t}$ or $\PHIm{\beta}{t}$ compactly supported,
generally, nor are they bounded on the same side.  So we need to
make sense of the convolution in (\ref{dder}).  To give meaning to
the convolution, let us note that the Fourier transform of
$\PHIp{\beta}{t}$ is given by \cite{GeSh64}
\begin{eqnarray}\label{PHIpFourier}
\PHIp{\beta}{t} & \stackrel{\mathcal{F}}{\longleftrightarrow} &
        \frac{\exp[\mathrm{sgn}(\omega)i\beta \pi/2]}{|\omega|^{\beta}}
\end{eqnarray}
and for $\PHIm{\beta}{t}$
\begin{eqnarray}\label{PHImFourier}
\PHIm{\beta}{t} & \stackrel{\mathcal{F}}{\longleftrightarrow} &
        \frac{\exp[-\mathrm{sgn}(\omega)i\beta \pi/2]}{|\omega|^{\beta}}
\end{eqnarray}
[Note that (\ref{PHIpFourier}) and (\ref{PHImFourier}) imply that,
up to a sign, the fractional derivatives go to the integer ordered
derivatives when $\beta$ is an integer.]  Then,
\begin{eqnarray}\label{PHIpmconv}
\PHIm{\beta}{t} \ast \PHIp{\beta}{t} & \stackrel{\mathcal{F}}
        {\longleftrightarrow} & |\omega|^{-2\beta}
\end{eqnarray}
Now,
\begin{eqnarray}\label{PSI1}
\frac{|t|^{2\beta-1}}{2 \cos(\beta\pi) \Gamma(2\beta)}
    & \stackrel{\mathcal{F}}{\longleftrightarrow} &
    |\omega|^{-2\beta}
\end{eqnarray}
We will define
\begin{eqnarray}\label{PSIdef}
\PSI{2\beta}{t} & := & \PHIm{\beta}{t}\ast \PHIp{\beta}{t}
                        \nonumber       \\
            & = & \frac{|t|^{2\beta-1}}{2 \cos(\beta\pi) \Gamma(2\beta)}
\end{eqnarray}
and let
\begin{eqnarray}\label{FFD1}
\PHIm{\beta}{t}\ast \PHIp{\beta}{t}\ast f(t) & \equiv &
            \PSI{2\beta}{t}\ast f(t)
\end{eqnarray}
for any $\beta$ where $f(t) \equiv 0$ for $t<a$ and $t>b$.  We
call (\ref{FFD1}) a Feller fractional derivative [FFD]
\cite{Podl01} and write this as
\begin{eqnarray}\label{FFDdef}
\FFD{a}{b}{2\alpha}{f} & := & \PSI{-2\alpha}{t} \ast f(t)     \\
        & = & \RFD{b}{\alpha}{\LFD{a}{\alpha}{f}}   \nonumber
\end{eqnarray}
Note that, for $n$ an integer,
\begin{eqnarray}
\FFD{a}{b}{2n}{f} & = & (-1)^{n}f^{(2n)}(t)  \label{Feller2n}
\end{eqnarray}
for\ $0<t<T$, but
\begin{eqnarray}\label{Feller2n+1}
\FFD{a}{b}{2n+1}{f} & \neq & \pm f^{(2n+1)}(t)
\end{eqnarray}

Some care is needed when using the FFDs.  Formally we have set
$f(t) \equiv 0$ for $t>a$ and $t<b$.  However, the LFD only acts
on the resulting discontinuities that may be present in $f(t)$ at
$t = a$, not at $t=b$.  Conversely, the RFD acts on the
discontinuities at $t=b$, not $t=a$.  It is perhaps better to
write (\ref{PSIdef}) as
\begin{eqnarray}\label{Psiredef}
\PSI{2\beta}{t} & = & \frac{1}{2 \cos(\beta\pi)}\ \left[
        \PHIp{2\beta}{t} + \PHIm{2\beta}{t}\right]
\end{eqnarray}
Then (\ref{FFDdef}) can be written as
\begin{eqnarray}\label{FFDredef}
\FFD{a}{b}{2\alpha}{f} & = & \frac{1}{2\cos(\beta\pi)}\ \left\{
        \LFD{a}{2\alpha}{f}
        + \RFD{b}{2\alpha}{f} \right\}
\end{eqnarray}
We will take (\ref{FFDdef}) as implying (\ref{FFDredef}).

In general, the fractional derivatives are nonlocal in time.  That
is, they have a ``memory''.  For integer ordered LFDs and RFDs,
this memory disappears [i.e., they are ``amnesiac''] and they act
locally in time. Even integer ordered FFDs are also amnesiac since
the kernels $\PSI{-2n}{t}$ equal, up to a sign, $\PHIp{-2n}{t}$
and $\PHIm{-2n}{t}$ in this case.  All of the fractional
derivatives have a fading memory, however \cite{BoCh85}.  That is,
they are affected more by the recent past and/or future than the
distant past and/or future.

\section{Riewe's Fractional Lagrangian Mechanics}
\label{SEC-Riewe}

Here we examine Riewe's fractional mechanics \cite{Riew96,Riew97},
restricting our attention to Lagrangian mechanics with Lagrangians
of the form
\begin{eqnarray}\label{RFMLag1}
L(q,\lfq{a}{\alpha},\lfq{a}{1}) & = & \frac{m}{2}\
    \left(\lfq{a}{1}\right)^{2} +
    \frac{C}{2}\left(\lfq{a}{\alpha}\right)^{2} - V(q)
\end{eqnarray}
where $q$ is our [generalized] coordinate, $C$ is a constant,
$0<\alpha<1$ and,
\begin{eqnarray}\label{lfgdef}
\lfq{a}{\alpha}& := & \LFD{a}{\alpha}{q}
\end{eqnarray}
We define the action associated with (\ref{RFMLag1}) by
\begin{eqnarray}\label{Raction1}
S[q] & := & \int_{a}^{b} L dt
\end{eqnarray}
Let us consider perturbations $\eta(t)$ of $q(t)$ where $\eta(t)$
vanishes for $t \leq a$ and $t \geq b$ but is otherwise arbitrary.
Then,
\begin{eqnarray}\label{Rpert1}
\delta S[q] & = & \delta \int_{a}^{b} L dt  \nonumber   \\
        & = & \int_{a}^{b} \left[ L(q+\eta,\lfq{a}{\alpha} +
        \lfeta{a}{\alpha},\lfq{a}{1} + \lfeta{a}{1}) - L(q,\lfq{a}{\alpha},
        \lfq{a}{1})\right]dt
\end{eqnarray}
Expanding the perturbed Lagrangian in (\ref{Rpert1})
\begin{eqnarray}\label{Rpert2}
L(q+\eta,\lfq{a}{\alpha} + \lfeta{a}{\alpha},\lfq{a}{1} +
    \lfeta{a}{1})
    & = & L(q,\lfq{a}{\alpha},\lfq{a}{1}) +   \nonumber   \\
    & &  \frac{\partial L}{\partial q}\ \eta + \frac{\partial L}{\partial
    \lfq{a}{\alpha}}\ \lfeta{a}{\alpha} + \frac{\partial L}{\partial
    \lfq{a}{1}}\ \lfeta{a}{1}
\end{eqnarray}
and using (\ref{Rpert2}) in (\ref{Rpert1}), we have
\begin{eqnarray}\label{Rpert3}
\delta S[q] & = & \int_{a}^{b} \left\{\frac{\partial
    L}{\partial
    q}\ \eta + \frac{\partial L}{\partial \lfq{a}{\alpha}}\
    \lfeta{a}{\alpha} + \frac{\partial L}{\partial \lfq{a}{1}}\
    \lfeta{a}{1}\right\} dt \nonumber       \\
    & = & \int_{a}^{b} \eta \left\{\frac{\partial L}{\partial q} +
    \RFD{b}{\alpha}{\frac{\partial L}{\partial \lfq{a}{\alpha}}} +
    \RFD{b}{1}{\frac{\partial L}{\partial \lfq{a}{1}}}\right\} dt
\end{eqnarray}
where we used (\ref{IBP}) in going to the second equality.

Hamilton's principle states that the actual path that a system
follows will be that which causes (\ref{Rpert3}) to vanish. Since
$\eta$ is infinitesimal but arbitrary, the bracketed term in
(\ref{Rpert3}) must vanish for $\delta S[q]$ to vanish.  Hence,
our {Euler-Lagrange equation} is
\begin{eqnarray}\label{RELeq1}
\RFD{b}{1}{\frac{\partial L}{\partial \lfq{a}{1}}} +
        \RFD{b}{\alpha}{\frac{\partial L}{\partial \lfq{a}{\alpha}}} & =
        & - \frac{\partial L}{\partial q}
\end{eqnarray}
For our Lagrangian in (\ref{RFMLag1}), we have the following
Euler-Lagrange equation of motion
\begin{eqnarray}\label{RELeq2}
\RFD{b}{1}{m\lfq{a}{1}} +
        \RFD{b}{\alpha}{C\lfq{a}{\alpha}}
        & = & m \FFD{a}{b}{2}{q}  + C \FFD{a}{b}{2\alpha}{q}    \\
        & = & \frac{\partial V}{\partial q}    \nonumber
\end{eqnarray}
[From (\ref{FFDredef}), we see that (\ref{RELeq2}) is a
two-endpoint equation \cite{BiRo89}.]  If, for example, $V(q) =
1/2 m \omega^{2} q^{2}$, (\ref{RELeq2}) can be written as
\begin{eqnarray}\label{HO1}
\left[m \Psi_{-2} + C\Psi_{-2\alpha} - m \omega^{2} \Psi_{0}
        \right]\ast q & = & 0
\end{eqnarray}

Notice the appearance of the FFD in (\ref{RELeq2}).  It arises
because of the integration by parts formula (\ref{IBP}).  In order
to have a strictly causal equation of motion, Riewe suggests
considering an infinitesimal time interval $[0,2\epsilon]$ and
then replacing all RFDs with LFDs. This seems unsatisfactory
because fractional operators have memory due to their nonlocal [in
time] nature.  By restricting the time interval to an
infinitesimal duration, Riewe is effectively erasing this memory.
Also, it is questionable if this will provide an accurate
approximation.  For example, let our time period be
$[0,2\epsilon]$ and
\begin{eqnarray}\label{fexample}
f(t) & = & \delta(t-\epsilon)   \\
    & = & \PHIp{0}{t-\epsilon}  \nonumber
\end{eqnarray}
Then,
\begin{eqnarray}\label{PHIexample}
\LFD{a}{2\alpha}{f} & = & \PHIp{-2\alpha}{t-\epsilon}
\end{eqnarray}
but,
\begin{eqnarray}\label{PSIexample}
\FFD{a}{b}{2\alpha}{f} & = & \PSI{-2\alpha}{t-\epsilon}
\end{eqnarray}
Now let $\alpha = 1/2$.  Obviously (\ref{PHIexample}) and
(\ref{PSIexample}) do not agree for $t<\epsilon$.  For $t >
\epsilon$ we have that $\PSI{-1}{t-\epsilon}  \neq  0$ while
$\PHIp{-1}{t-\epsilon}$ does vanish.

If we blindly follow the above procedure for (\ref{RELeq2}) we
have the resulting equation
\begin{eqnarray}\label{RELeqex}
m \ddot{q} + C \dot{q} = \frac{\partial V}{\partial q}
\end{eqnarray}
which is missing a minus sign in front of the derivative of the
potential $V$.  We could of course recognize that
$\FFD{a}{b}{2}{q} = -\ddot{q}$ for $a<t<b$ and change the sign of
$C$ in (\ref{RFMLag1}).  Then we would have the correct causal
equation of motion with friction
\begin{eqnarray}\label{RELeqex2}
m \ddot{q} + C \dot{q} = - \frac{\partial V}{\partial q}
\end{eqnarray}
However, this requires that we treat integer ordered derivatives
differently, which is not entirely satisfactory.

Instead of using the Lagrangian in (\ref{RFMLag1}), let us use
\begin{eqnarray}\label{RFMLag2}
L & = & -\frac{m}{2}\ \left(\LFD{a}{1}{q}\right)
    \left(\RFD{b}{1}{q}\right) - \frac{C}{2}\
    \left(\LFD{a}{\alpha}{q}\right)
    \left(\RFD{b}{\alpha}{q}\right) - V(q)
\end{eqnarray}
If we perturb $q$ by $\eta$ in (\ref{RFMLag2}), we have, to first
order in $\eta$,
\begin{eqnarray}\label{pertLag}
\delta L & = & -\frac{m}{2}\ \LFD{a}{1}{q} \rfeta{b}{1}
        -\frac{C}{2} \LFD{a}{\alpha}{q} \rfeta{b}{\alpha}
        -\frac{1}{2}\ \frac{\partial V}{\partial q}\ \eta \nonumber   \\
        & & -\frac{m}{2}\ \RFD{b}{1}{q} \lfeta{a}{1}
        -\frac{C}{2} \RFD{b}{\alpha}{q} \lfeta{a}{\alpha}
        -\frac{1}{2}\ \frac{\partial V}{\partial q}\ \eta
\end{eqnarray}
Then, using (\ref{IBP}),
\begin{eqnarray}\label{pertLag2}
\int_{a}^{b} \delta L\ dt & = & \underbrace{\int_{a}^{b}
    \frac{\eta}{2}\ \left\{-m \LFD{a}{2}{q}
        - C
    \LFD{a}{2\alpha}{q} - \frac{\partial V}{\partial q}\
    \right\}dt}_{\mathrm{retarded}}   +
    \nonumber       \\
    & & \underbrace{\int_{a}^{b} \frac{\eta}{2}\ \left\{-m \RFD{b}{2}{q} - C
    \RFD{b}{2\alpha}{q} - \frac{\partial V}{\partial q}\
    \right\}dt}_{\mathrm{advanced}}
\end{eqnarray}
Now,
\begin{eqnarray}\label{pertS1}
\delta S[q] & = & \int_{a}^{b} \delta L dt
\end{eqnarray}
To make $\delta S[q]$ vanish, we will require that the bracketed
terms in (\ref{pertLag2}) vanish separately.  This gives us two
equations of motion
\begin{eqnarray}
\begin{array}{lr}
m \LFD{a}{2}{q} + C \LFD{a}{2\alpha}{q} = -\frac{\partial
        V}{\partial q}&
        \mathrm{(retarded)}     \label{retardedeom1}
\end{array}\\
\begin{array}{lr}
m \RFD{b}{2}{q} + C \RFD{b}{2\alpha}{q} = -\frac{\partial
        V}{\partial q} &
        \mathrm{(advanced)}     \label{advencedeom1}
\end{array}
\end{eqnarray}
For the special case $\alpha = 1/2$, (\ref{retardedeom1}) and
(\ref{advencedeom1}) become
\begin{eqnarray}
\begin{array}{lr}
m \ddot{q} + C \dot{q} = -\frac{\partial
        V}{\partial q}&
        \mathrm{(retarded)}     \label{retardedeom2}
\end{array}\\
\begin{array}{lr}
m \ddot{q} - C\dot{q} = -\frac{\partial
        V}{\partial q} &
        \mathrm{(advanced)}     \label{advencedeom2}
\end{array}
\end{eqnarray}
respectively, for $a < t < b$.

Comparing (\ref{retardedeom2}) and (\ref{advencedeom2}) with
(\ref{Intro2}), we see that Bateman's method is included in
Riewe's formalism provided we use Lagrangians as in
(\ref{RFMLag2}) and, require the advanced and retarded parts of
the perturbed action to vanish separately.  [These types of
Lagrangians were not considered explicitly by Riewe in
\cite{Riew96,Riew97}.]  Allowing both a retarded and an advanced
equation of motion to arise from the variation of the action seems
more natural than, for example, (\ref{HO1}).  It avoids the final
procedure of replacing $\RFD{b}{\alpha}{q}$ with
$\LFD{a}{\alpha}{q}$.  Also, the Lagrangian in (\ref{RFMLag2}) is
preferable to that in (\ref{RFMLag1}) because it does not apriori
assume that the LFDs are to be favored over the RFDs.  Now we turn
our attention to an alternate way of constructing nonconservative
Lagrangians.

\section{Volterra Series}
\label{SEC-Volterra}

In order to develop our new formalism of nonconservative
Lagrangians, we will need some background on Volterra series
\cite{BCD84,Stev95}. The Volterra series is a generalization to
functionals of the power series of a function. For some functional
$\mathcal{V}[q]$, we define the symmetric kernels
\begin{eqnarray}\label{Ks}
\K{n}{s}{\tau_{1},\ldots,\tau_{n}} & := & \frac{\delta^{n}
\mathcal{V}[q]}
        {\delta q(\tau_{1}) \cdots \delta q(\tau_{n})}
\end{eqnarray}
The $\K{n}{s}{\cdot}$'s are symmetric under an interchange of the
$\tau_{i}$'s.  So, for example, $\K{2}{s}{\tau_{1},\tau_{2}} =
\K{2}{s}{\tau_{2},\tau_{1}}$.  Introducing the notation
\begin{eqnarray}\label{Knstar}
\Kstar{n}{s}{n} & := & \int_{\tau_{1}} \cdots \int_{\tau_{n}}
        \K{n}{s}{\tau_{1},\ldots,\tau_{n}} q(\tau_{n}) \cdots
        q(\tau_{1}) d\tau_{n} \cdots d\tau_{1}
\end{eqnarray}
we can expand the functional $\mathcal{V}[q]$ in the Volterra
series
\begin{eqnarray}\label{VolterraSeries}
\mathcal{V}[q] & = & \mysum \frac{1}{n!} \Kstar{n}{s}{n}
\end{eqnarray}
[For our purposes we can assume that $K_{0}^{(s)} = \mathcal{V}[0]
\equiv 0$.]  It is easy to show that
\begin{eqnarray}\label{DiffKstar}
\frac{\delta \Kstar{n}{s}{n}}{\delta q(t)} & = & n
        \Kstar{n}{s}{n-1}       \nonumber       \\
        & := & n \int_{\tau_{2}} \cdots \int_{\tau_{n}}
        \K{n}{s}{t,\tau_{2},\ldots,\tau_{n}}  \\
        & & \hspace{.75in} q(\tau_{n}) \cdots
        q(\tau_{2}) d\tau_{n} \cdots d\tau_{2}  \nonumber
\end{eqnarray}

The symmetric kernels are the natural choice to use in a Volterra
series.  However, we may be given asymmetric kernels and would
like to symmetrize them or vice versa.  As motivation, consider
the function
\begin{eqnarray}\label{qex1}
v(\mathbf{q}) & = & \frac{1}{2}\ \sum_{i} K_{ii} q_{i}^{2} +
        \sum_{i < j} K_{ij}q_{i} q_{j}
\end{eqnarray}
where $\mathbf{q} = [q_{1},\ldots, q_{n}]$.  We can symmetrize
$v(\mathbf{q})$ into the form
\begin{eqnarray}\label{gex2}
v(\mathbf{q}) & = & \frac{1}{2}\ \sum_{i,j} K_{ij} q_{i}q_{j}
\end{eqnarray}
where $K_{ij} = K_{ji}$ is a symmetric matrix.  We will be
particularly interested in triangular kernels given by
\begin{eqnarray}\label{triK}
\K{n}{t}{\tau_{1},\ldots,\tau_{n}} & = & 0 \hspace{.5in}
        \mbox{unless $\tau_{1}
        \geq \tau_{2} \geq \cdots \geq \tau_{n}$}
\end{eqnarray}
Now, let $\sigma$ be a permutation of $1,\ldots,n$.  The
symmetrization of (\ref{triK}) is defined as
\begin{eqnarray}\label{symK}
\mathrm{sym} \K{n}{t}{\tau_{1},\ldots,\tau_{n}} & := &
        \frac{1}{n!}\ \sum_{\sigma} \K{n}{t}{\tau_{\sigma_{1}},\ldots,
        \tau_{\sigma_{n}}}      \\
        & = & \frac{1}{n!}\ \K{n}{s}{\tau_{1},\ldots,\tau_{n}}
        \nonumber
\end{eqnarray}

\section{Volterra Series Fractional Lagrangian Mechanics}
\label{SEC-FRACMECH}

Let us now reconsider the nonconservative harmonic oscillator
equation of motion in (\ref{HO1}).  Using the notation in
(\ref{DiffKstar}), (\ref{HO1}) becomes
\begin{eqnarray}\label{HOKstar}
\Kstar{2}{s}{1} & = & 0
\end{eqnarray}
where
\begin{eqnarray}\label{K2s1}
\K{2}{s}{t,\tau} & := & m\PSI{-2}{t-\tau} +
        C\PSI{-2\alpha}{t-\tau} -m\omega^{2} \PSI{0}{t-\tau}
\end{eqnarray}
Let our action be given by
\begin{eqnarray}\label{VSaction1}
\mathcal{V}_{2}[q] & = & \frac{1}{2}\ K_{2}^{(s)} \star q^{2}
\end{eqnarray}
Then,
\begin{eqnarray}\label{EOMVS1}
\frac{\delta \mathcal{V}_{2}[q]}{\delta q(t)} & = &
        \Kstar{2}{s}{1}
\end{eqnarray}
Requiring (\ref{EOMVS1}) to vanish gives us (\ref{HOKstar}).

Suppose now that we have a driven harmonic oscillator
\begin{eqnarray}\label{DrivenHO}
m\ddot{q} + m\omega^{2} q & = & f(t)
\end{eqnarray}
We can form a new functional
\begin{eqnarray}\label{DHOG1}
\mathcal{V}_{2^{'}}[q] & = &  \Kstar{1^{'}}{s}{1} + \frac{1}{2}\
        \Kstar{2^{'}}{s}{2}
\end{eqnarray}
where
\begin{eqnarray}\label{K2prime}
\K{2^{'}}{s}{t,\tau} & := & m\PSI{-2}{t-\tau}  -m\omega^{2}
        \PSI{0}{t-\tau}
\end{eqnarray}
It immediately follows that, ignoring boundary conditions,
\begin{eqnarray}\label{DelG1}
\frac{\delta \mathcal{V}_{2^{'}}[q]}{\delta q(t)} & = &
        \K{1^{'}}{s}{t} -
        m\ddot{q}-m\omega^{2} q
\end{eqnarray}
Requiring (\ref{DelG1}) to vanish and comparing with
(\ref{DrivenHO}), we see that $\K{1^{'}}{s}{t} = f(t)$.  We can
also handle higher order potentials.  Let, for example,
\begin{eqnarray}\label{G3}
\mathcal{V}_{3^{'}}[q] & = & \Kstar{1^{'}}{s}{1} + \frac{1}{2!}\
        \Kstar{2^{'}}{s}{2} + \frac{1}{3!}\ \Kstar{3^{'}}{s}{3}
\end{eqnarray}
where, for some constant $C$,
\begin{eqnarray}\label{K3}
K_{3^{'}}^{(s)}(\tau_{1},\tau_{2},\tau_{3}) & := & C
        \PSI{0}{\tau_{1} -\tau_{2}}\PSI{0}{\tau_{2} - \tau_{3}}
\end{eqnarray}
Then, again ignoring boundary terms,
\begin{eqnarray}\label{DelG3}
\frac{\delta \mathcal{V}_{3^{'}}[q]}{\delta q(t)} & = & f(t) -
        m\ddot{q} - m\omega^{2} q + \frac{C}{2}\ q^{2}
\end{eqnarray}
We recognize (\ref{G3}) as the beginning of the Volterra series
for some functional $\mathcal{V}[q]$.  To all orders of $q$,
\begin{eqnarray}\label{GenAction}
\mathcal{V}[q] & = & \mysum \frac{1}{n!}\ \Kstar{n}{s}{n}
\end{eqnarray}
[We can ignore the $n=0$ term in (\ref{GenAction}) since this only
adds an irrelevant constant to $\mathcal{V}[q]$.]  For $n \geq 2$,
the $\K{n}{s}{\cdot}$'s are interpreted as the environment's
reaction to $q$, which affects $q$'s evolution.  Any forcing
function is included in $\K{1}{s}{t}$.

All of the actions considered above share two key properties:
\begin{enumerate}
    \item\label{P1} The kernels $\K{n}{s}{\cdot}$ are all localized
    along the line $\tau_{1} = \tau_{2}$.
    \item\label{P2} The kernels satisfy the relation
    $\K{n}{s}{\cdot} = \K{n}{t}{\cdot}$.
\end{enumerate}
These properties make the above actions particularly easy to
analyze. However, it is impossible to introduce even the simple
term $C\dot{q}$ into the equations of motion using the
$\Psi_{\alpha}$'s [see (\ref{Feller2n+1})]. Using triangular,
instead of symmetric, kernels results in a more flexible
formalism.  This amounts to using the $\Phi_{\alpha}^{\pm}$'s in
the Volterra series instead of the $\Psi_{\alpha}$'s.  We will
then be able to construct symmetric kernels that only use the
$\Phi^{\pm}_{\alpha}$'s, not the $\Psi_{\alpha}$'s.  This requires
us to be careful about the boundary terms in our equations. It is
this situation that we now turn our attention to.

We return again to the nonconservative harmonic oscillator.  For
some constant $C$, define the triangular kernels
\begin{eqnarray}
K^{+}_{2}(t,\tau) & := & -\left[m\PHIp{-2}{t-\tau} +
        C\PHIp{-2\alpha}{t-\tau} + m\omega^{2}
        \PHIp{0}{t-\tau}\right]
        \label{Kp2}             \\
K^{-}_{2}(\tau,t) & := & -\left[m\PHIm{-2}{\tau -t} +
        C\PHIm{-2\alpha}{\tau-t} + m\omega^{2}
        \PHIm{0}{\tau-t}\right]
        \label{Km2}
\end{eqnarray}
where
\begin{eqnarray}\label{Kp2=Km2}
K^{+}_{2}(t,\tau) & = & K^{-}_{2}(\tau,t)
\end{eqnarray}
Now consider the functional
\begin{eqnarray}\label{Gplus}
\widehat{\mathcal{V}}[q] & := & \frac{1}{2}\ \int_{a^{-}}^{b^{+}}
        \int_{a^{-}}^{\tau_{1}^{-}}
        K_{2}^{+}
        (\tau_{1},\tau_{2}) q(\tau_{2}) q(\tau_{1}) d\tau_{2}
        d\tau_{1}
\end{eqnarray}
The functional derivative of (\ref{Gplus}) is given by
\cite{Stev95}
\begin{eqnarray}\label{derGplus}
\frac{\delta \widehat{\mathcal{V}}[q]}{\delta q(t)} & = & \lim_{h
        \rightarrow 0}
        \frac{1}{2h}\ \left\{ \int_{a^{-}}^{b^{+}} \int_{a^{-}}^{\tau_{1}^{-}}
        K^{+}_{2}(\tau_{1},\tau_{2}) [q(\tau_{2}) + h
        \delta(\tau_{2} -t)]    \times  \nonumber \right.  \\
        & & \hspace{1.3in} [q(\tau_{1}) + h \delta( \tau_{1} - t)]
        d\tau_{2} d\tau_{1}  -   \nonumber           \\
        & & \left. \hspace{.5in}  \int_{a^{-}}^{b^{+}} \int_{a^{-}}^{\tau_{1}^{-}}
        K^{+}_{2}(\tau_{1},\tau_{2}) q(\tau_{2}) q(\tau_{1})
        d\tau_{2} d\tau_{1} \right\}    \nonumber           \\
        & = & \frac{1}{2}\ \int_{a^{-}}^{b^{+}} \int_{a^{-}}^{\tau_{1}^{-}}
        K^{+}_{2}(\tau_{1},\tau_{2}) \delta(\tau_{2} -t)
        q(\tau_{1}) d\tau_{2} d\tau_{1}  +   \nonumber       \\
        & & \frac{1}{2}\ \int_{a^{-}}^{b^{+}} \int_{a^{-}}^{\tau_{1}^{-}}
        K^{+}_{2}(\tau_{1},\tau_{2}) q(\tau_{2}) \delta(\tau_{1} -
        t) d\tau_{2} d\tau_{1}      \nonumber       \\
        & = & \underbrace{\frac{1}{2}\ \int_{t^{+}}^{b^{+}} K^{-}_{2}(t,\tau_{1}) q(\tau_{1})
        d\tau_{1}}_{\mathrm{advanced}} + \underbrace{\frac{1}{2}\ \int_{a^{-}}^{t^{-}} K^{+}_{2}(t,\tau_{2})
        q(\tau_{2})d\tau_{2}}_{\mathrm{retarded}}
\end{eqnarray}
where $a \leq t \leq b$.  Instead of requiring the sum in
(\ref{derGplus}) to vanish, we will require the advanced and
retarded parts of the action's variation to vanish separately.
This gives us two equations of motion for our system
\begin{eqnarray}
\begin{array}{lr}
\left[m\PHIp{-2}{t} + C\PHIp{-2\alpha}{t} + m\omega^{2}
        \PHIp{0}{t}\right]\ast q(t)  =  0 &
        \mathrm{(retarded)}     \label{retardedeom}
\end{array}\\
\begin{array}{lr}
\left[m\PHIm{-2}{t} +C\PHIm{-2\alpha}{t} + m\omega^{2}
        \PHIm{0}{t}\right]\ast q(t)  =  0  &
        \mathrm{(advanced)}     \label{advancedeom}
\end{array}
\end{eqnarray}
From (\ref{derGplus}), we see that $q(\tau_{1}) \equiv 0$ for
$\tau_{1}>b$ and $q(\tau_{2}) \equiv 0$ for $\tau_{2}<a$ in
(\ref{Gplus}).

Note that if our kernels only contain terms $\Phi_{2n}^{\pm}$, $n$
an integer, requiring the advanced and retarded parts to vanish
separately is equivalent to requiring the sum in (\ref{derGplus})
to vanish, ignoring boundary conditions.  This is because
$\Phi_{2n}^{+} = \Phi_{2n}^{-}$ and both equal, up to a sign,
$\Psi_{2n}$.  So in this case we can freely use the symmetric
kernels $\Psi_{2n}$ in our action.  We can also extend the above
action to a driven harmonic oscillator and higher order
potentials, as we did earlier.  Again, this is due to the fact
that $\Psi_{0} = \Phi_{0}^{\pm}$ and also that $K_{1}^{(s)} =
K_{1}^{\pm}$.

The kernel in (\ref{Gplus}) is lower triangular in the
$\tau_{1}\tau_{2}$-plane [i.e., $K^{+}_{2}(\tau_{1},\tau_{2})
\equiv 0$ when $\tau_{1} \leq \tau_{2}$].  We could have equally
well used the functional
\begin{eqnarray}\label{Gplus2}
\widetilde{\mathcal{V}}[q] & := & \frac{1}{2}\
        \int_{a^{-}}^{b^{+}}
        \int_{\tau_{1}^{+}}^{b^{+}}
        K_{2}^{+}
        (\tau_{2},\tau_{1}) q(\tau_{2}) q(\tau_{1}) d\tau_{2}
        d\tau_{1}
\end{eqnarray}
to arrive at the equations of motion in (\ref{retardedeom}) and
(\ref{advancedeom}).  Here the kernel is upper triangular in the
$\tau_{1}\tau_{2}$-plane [i.e., $K^{+}_{2}(\tau_{2},\tau_{1})
\equiv 0$ when $\tau_{1} \geq \tau_{2}$]. A derivation similar to
that in (\ref{derGplus}) shows that, if we use (\ref{Gplus2}) for
our action , then $q(\tau_{1}) \equiv 0$ for $\tau_{1}<a$ and
$q(\tau_{2}) \equiv 0$ for $\tau_{2}>b$.  It follows that the
symmetric action
\begin{eqnarray}\label{Gplussym}
\mathcal{V}[q] & := & \frac{1}{2}\ \left\{\widehat{\mathcal{V}}
        [q] +
        \widetilde{\mathcal{V}}[q] \right\}       \\
        & = &\frac{1}{2}\  \int_{a^{-}}^{b^{+}}
        \int_{a^{-}}^{b^{+}} \left\{\frac{1}{2}\
        \left[K_{2}^{+}(\tau_{1},\tau_{2}) +
        K_{2}^{+}(\tau_{2},\tau_{1}) \right]\right\}
         q(\tau_{2}) q(\tau_{1}) d\tau_{2}
        d\tau_{1}               \nonumber
\end{eqnarray}
could also be used to derive (\ref{retardedeom}) and
(\ref{advancedeom}), where $q(\tau_{i}) \equiv 0$, $i=1,2$, for
$\tau_{i} < a$ and $\tau_{i} > b$.  The above is easier to see if
we let $K^{+}_{2}(t) := K^{+}_{2}(t,0)$ and $K^{-}_{2}(t) :=
K^{-}_{2}(t,0)$.  Then (\ref{Gplus}) is given by
\begin{eqnarray}\label{Gplusconv}
\widehat{\mathcal{V}}[q] & = & \frac{1}{2}\ \int q(t) \left[
            K^{+}_{2}(t) \ast q(t)
            \right] dt
\end{eqnarray}
Using the integration by parts formula in (\ref{IBP2}) gives us
(\ref{Gplus2})
\begin{eqnarray}\label{Gtildeconv}
\widetilde{\mathcal{V}}[q] & = &\frac{1}{2}\  \int q(t) \left[
        K^{-}_{2}(t) \ast q(t)
        \right] dt
\end{eqnarray}
Adding (\ref{Gplusconv}) to (\ref{Gtildeconv}), and multiplying by
$1/2$, results in (\ref{Gplussym})
\begin{eqnarray}\label{Gsymconv}
\mathcal{V}[q] & = &\frac{1}{2}\  \int q(t) \left\{ \frac{1}{2}\
        \left[ K^{+}_{2}(t) +
        K^{-}_{2}(t) \right] \ast q(t) \right\} dt
\end{eqnarray}

Let us now collect some remaining observations.  The usual action
for the harmonic oscillator is given by
\begin{eqnarray}\label{ex1}
S[q] & = & \frac{1}{2}\ \int \left[ m \dot{q}^{2} - m \omega^{2}
        q^{2} \right] dt        \nonumber       \\
        & = & - \frac{1}{2}\ \int q \left[ m \ddot{q} + m
        \omega^{2} q \right] dt + \left. \frac{1}{2}\ q \dot{q}
        \right|_{a}^{b}
\end{eqnarray}
where we used an integration by parts in the second equality.  The
Volterra series action in (\ref{Gplusconv}), with $C = 0$ in
(\ref{Kp2}), gives
\begin{eqnarray}\label{ex2}
\widehat{\mathcal{V}}[q] & = & - \frac{1}{2}\ \int q \left[ m
        \Phi^{+}_{-2} + m \omega^{2} \Phi^{+}_{0} \right] \ast q\
        dt                  \nonumber           \\
        & = & - \frac{1}{2} \int q \left[ m \ddot{q} + m
        \omega^{2} q \right] dt - \frac{1}{2}\ \int q \left[
        \dot{q}(a) \delta(t-a) + q(a) \dot{\delta}(t-a) \right] dt
        \nonumber                                               \\
        & = & -\frac{1}{2}\ \int q \left[ m \ddot{q} + m
        \omega^{2} q \right] dt
\end{eqnarray}
where we used (\ref{CGrel}) for the second equality.  Hence,
\begin{eqnarray}\label{ex3}
S[q] & = & \widehat{\mathcal{V}}[q] + \left. \frac{1}{2}\ q\dot{q}
        \right|_{a}^{b}
\end{eqnarray}
Thus, $\widehat{\mathcal{V}}[q]$ differs from $S[q]$ only by the
boundary terms of $q(t)$, which, by our above analysis, does not
affect the resulting equation of motion. This also holds for
(\ref{Gtildeconv}) and (\ref{Gsymconv}).  Returning to
(\ref{Gplusconv}), let us perturb $q(t)$ by $\eta(t)$.  Then,
\begin{eqnarray}\label{ex4}
\delta \widehat{\mathcal{V}}[q] & = & \underbrace{\frac{1}{2}\
            \int q \left[
            K^{+}_{2} \ast \eta \right] dt}_{\mathrm{advanced}} +
            \underbrace{ \frac{1}{2}\ \int \eta \left[ K^{+}_{2}
            \ast q \right] dt}_{\mathrm{retarded}}
\end{eqnarray}
The second term on the right of (\ref{ex4}) is what we typically
want in order to derive our equation of motion for $q(t)$.
However, the first term on the right of (\ref{ex4}) is
interesting.  It shows that the advanced equation of motion for
$q(t)$ arises because of the perturbation of the environment's
reaction due to $\eta(t)$. That is, using (\ref{IBP2}),
\begin{eqnarray}\label{ex5}
\int q \left[ K^{+}_{2} \ast \eta \right] dt & = & \int \eta
        \left[ K^{-}_{2} \ast q \right] dt
\end{eqnarray}
So it seems that the future evolution of $q(t)$ is affected by its
past evolution because of the memory ``stored'' in the
environment.

\section{Perturbing the Environment}
\label{SEC-ENVIR}

So far we have examined everything in a fairly standard way.  We
assumed that the environment is described by $K^{+}_{2}(t) :=
K^{+}_{2}(t,0)$ [see (\ref{Kp2})] and introduced a particle into
this environment via $q(t)$.  Then we perturbed the particle's
path and required that the variation in the action vanish under
this perturbation. Notice that the particle did not change the
environment's kernel given by $K^{+}_{2}(t)$.  So the particle
itself must be so negligible that the environment's kernel does
not substantially change under its introduction.  That is, the
particle is a perturbation to the environment.  Let us explore
this idea more for the nonconservative harmonic oscillator.  This
will lead to a more holistic view of mechanics which ignores the
distinction between the environment and the system, in the case of
the harmonic oscillator.

Let us assume that the environment is adequately described by the
generalized function [see (\ref{phiadd}) for the second equality
below]
\begin{eqnarray}\label{ENVIRO1}
K^{+}_{2}(t) & := & m\PHIp{-2}{t} +
        C\PHIp{-2\alpha}{t} + m\omega^{2} \PHIp{0}{t}
                                   \nonumber          \\
        & = & \left[ m\PHIp{-2}{t} +
        C\PHIp{-2\alpha}{t} + m\omega^{2} \PHIp{0}{t}
        \right] \ast    \widehat{\Phi}_{0}^{+}(t)  \nonumber   \\
        & = & K^{+}_{2}(t) \ast \widehat{\Phi}_{0}^{+}(t)
\end{eqnarray}
where the hat on $\widehat{\Phi}_{0}^{+}(t)$ is for bookkeeping
purposes only.  Now let us perturb the environment $K^{+}_{2}(t)$
by perturbing $\widehat{\Phi}_{0}^{+}(t)$ by $\widehat{\eta}(t)$,
where $\widehat{\eta}(t)$ is infinitesimal in the, e.g.,
$L^{2}$-norm compared to $\widehat{\Phi}_{0}^{+}(t) =
\widehat{\delta}(t)$.  In particular, we will not require that
$\widehat{\eta}(t)$ vanish at any boundaries.  Then,
\begin{eqnarray}\label{ENVIRO2}
\delta K^{+}_{2}(t) & = & K^{+}_{2}(t) \ast
        \widehat{\eta}(t)
\end{eqnarray}
Requiring (\ref{ENVIRO2}) to vanish gives us exactly
(\ref{retardedeom}) when we identify $\widehat{\eta}(t) \equiv
q(t)$.

In (\ref{ENVIRO1}), we assumed that the environment reacts
causally to any perturbation.  This resulted in the retarded
equation of motion in (\ref{ENVIRO2}).  If, instead, we considered
the kernel $K^{-}_{2}(t) : = K^{-}_{2}(0,t)$, [see (\ref{Km2})],
then the advanced equation of motion in (\ref{advancedeom}) would
have resulted instead of (\ref{ENVIRO2}).  So, we see that it is
not necessary, for the nonconservative harmonic oscillator, to
construct an action in order to derive the equations of motion.
How far this idea can be advanced to more general systems is an
open question.

\section{Discussion}
\label{SEC-Discussion}

Let us look at the retarded equation in (\ref{retardedeom}) a
little.  This is a convolution between the coordinate $q(t)$ and
the distribution $K^{+}_{2}(t,0) := K^{+}_{2}(t)$.  An insightful
way of viewing this is to think of $K^{+}_{2}(t)$ as the
environment's response ``function'', where, for an arbitrary
$f(t)$,
\begin{eqnarray}\label{Disc1}
y(t) & = & K^{+}_{2}(t) \ast f(t)
\end{eqnarray}
Then, the actual paths that a system can follow will be those
$f(t)$ such that $y(t) \equiv 0$ in (\ref{Disc1}).  That is, the
path a system follows will be those such that the environment's
response to it vanishes.  This treatment of fractional derivatives
as signal processors is well known \cite{Rutm95} and can be
extended to more general Volterra series than that in
(\ref{retardedeom}) \cite{BoCh85,Pare70}.  Similar comments hold
for (\ref{advancedeom}) and its generalizations, but, in this
case, the environment's response is anti-causal.  We have not
pursued this line of research here.  However, it does open up the
possibility of examining mechanics from a systems theoretic
viewpoint [see, e.g., \cite{OpWi97} for an introduction to systems
theory].

We have not considered the most general action here.  Instead, our
attention was restricted to including fractional derivatives in an
equation of motion derived by using a variational principle.  In
this respect we have succeeded.  Further research is needed to see
how far our formalism can be developed and, how useful it will be
in situations other than those considered here.  In particular, it
would be interesting to extend the formalism in Section
\ref{SEC-ENVIR} to more general situations.

\section{Acknowledgements}
The authors would like to thank the NSF for grant \#9732986.


\bibliographystyle{plain}
\bibliography{FracDer}

\end{document}